# Optically driven ultrafast magnetic order transitions in two-dimensional ferrimagnets

Junjie He,[1,*] Thomas Frauenheim[1,2,*]

[1]*Bremen Center for Computational Materials Science, University of Bremen, Am Fallturm 1, 2835, Bremen, Germany.*

[2]*Beijing Computational Science Research Center (CSRC), Beijing 100193 and Shenzhen Computational Science and Applied Research (CSAR) Institute, Shenzhen 518110, China.*

E-mail: junjie.he.phy@gmail.com; thomas.frauenheim@bccms.uni-bremen.de

**ABSTRACT**

Laser-induced switching and manipulation of the spins in magnetic materials are of great interest to revolutionize future magnetic storage technology and spintronics with the fastest speed and least power dissipative. Inspired by the recent discovery of intrinsic two-dimensional (2D) magnets, which provide a unique platform to explore the new phenomenon for light-control magnetism in the 2D limit, we propose to realize light can efficiently tune magnetic states of 2D ferrimagnets in early time. Here, using the 2D ferrimagnetic MXenes ($M_2M'X_2F_2$, M/M'=Cr, V, Mo; X=C/N) as prototype systems, our real-time density functional theory (TDDFT) simulation show that laser pulses can directly induce ultrafast spin-selective charge transfer between two magnetic sublattices (M and M') on a few femtoseconds, and further generate dramatic changes in the magnetic structure of these MXenes, including a magnetic order transition from ferrimagnetism (FiM) to transient ferromagnetism (FM). The microscopic mechanism underpinning this ultrafast switching of magnetic order in MXenes is governed by optically induced inter-site spin transfer (OISTR) effect, which theoretically enables the ultrafast direct optical manipulation of the magnetic state in MXenes-based 2D materials. Our results open new opportunities for exploring the manipulation of the spin in 2D magnets by optical approaches.

## TOC

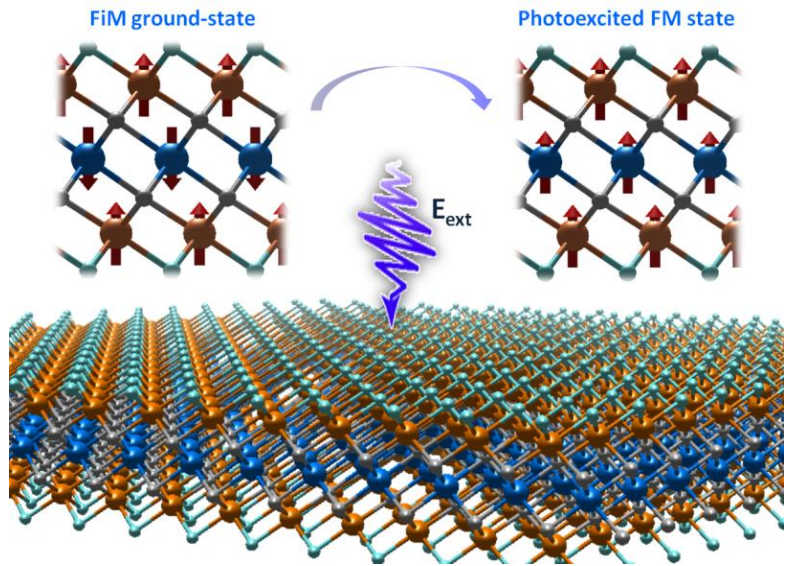

## Introduction

In 2017, two experimental groups have independently observed the intrinsic two-dimensional (2D) ferromagnetism (FM) in exfoliated van der Waals (vdW) $CrI_3$ and $CrGeTe_3$ crystal by magneto-optical technique. [1,2] Subsequently, a large variety of 2D magnetic vdW crystals, including $FePS_3$,[3] $NiPS_3$,[4] $MnPS_3$,[5] $Fe_3GeTe_2$,[6] $VSe_2$,[7] and $MnSe_2$[8] etc. have been realized by using only adhesive tape, chemical vapour deposition or molecular beam epitaxy. Interestingly, both $VSe_2$ and $MnSe_2$ were claimed to be a 2D ferromagnet with Curie temperature ($T_C$) above room temperature.[7,8] Until now, the family of 2D magnetic crystals is still rapidly growing, and many more new 2D magnets have been predicted and await discovery.

[9,10,11,12,13,14] These 2D magnets with excellent properties in optical, magnetic, magneto-electric, and magneto-optic areas, offer new opportunities for both exploring the manipulation of spin or magnetism in the 2D limit and for developing the next generation spintronic devices. Following these exciting discoveries, considerable studies have been devoted to manipulating spin interaction in 2D magnets, e.g., Dzyaloshinskii-Moriya interaction, interlayer magnetic order, and magnetic anisotropy by mechanical strain, gate voltage, and magnetic field.[15,16,17,18] For example, it demonstrated that the interlayer magnetic order of bilayer $CrI_3$ can be reversibly tuned from the antiferromagnetic (AFM) to the FM phase by applying electrostatic doping or vertical electric field gate voltage.[16,17,18]

The light represents the fastest means to manipulate the spin structures of matter at ultrashort time scales (from femtosecond to attosecond) that promises to revolutionize future recording and information processing by achieving the fastest possible and least power dissipative.[19] Laser induced spin injection in FM/NM interface,[20] all-optical magnetization switching,[21] optically tunable magnetic anisotropy in $CrI_3$ monolayer, [22] photo-induced topological states 2D materials,[23] and light can manipulate the interfacial magnetic proximity coupling and valley polarization $WSe_2/CrI_3$ heterostructures[24] are prominent examples of light-controlled properties in materials. Using time-dependent density-functional theory (TDDFT) simulation, Dewhurst $et\ al.$ have shown ultra-short laser can directly induce a spin transfer between sub-lattices causing significant magnetic order transition, i.e., the switching from Ferrimagnetic (FiM) to FM order in a multi-component magnetic system, which is governed by the previously unknown optically induced inter-site spin transfer (OISTR) effect in sub-exchange and sub-spin-orbit timescales.[25,26] Indeed, recently, several experimental groups have demonstrated the ultrafast laser can directly and coherently manipuate spin of magnetic multilayer and Heusler compounds on sub-femtosecond time scales, and induced spin transfer between two magnetic subsystems, which confirmed the OISTR effect.[19,27,28,29] These breakthroughs suggest a search for ultrafast optically tunable magnetism in 2D systems. However, to realize the OISTR effect, it is highly desirable to develop multi-component (two or more) magnetic systems but it is rare in 2D magnets.

Recently, 2D transition metal carbides or nitrides called by MXenes have emerged as a new type of 2D materials with the general formula $M_{n+1}X_nT_z$, where M represents an early transition metal (such as Ti, V, Cr, or Mo), X is carbon and/or nitrogen, and T stands for the surface terminations (e.g., -OH, -O, or -F).[30,31,32] The MXenes attract great attention due to their potential applications in sensors, catalysis, energy storage and nanoelectronics.[32] The chemical and physical properties of MXenes can be tuned through the choice of transition metals and surface chemical groups.[12,14,33,34,35] For example, the ferromagnetic behavior of bare $Cr_2C$ and $Cr_2TiC_2$ has been shown to become semiconducting antiferromagnetic upon -F, -OH and -Cl functionalization.[12,14] Particularly, the $Mo_3N_2F_2$ MXene has been found to be a ferrimagnetic half-metallicity with a high Curie temperature.[36] Therefore, MXenes represent unusual a class of multi-component magnetic materials in 2D system, which provide an excellent platform to study the optically tunable magnetism and spin transfer in 2D system.

In this work, we propose to realize the optically manipulate magnetic order transition in MXenes. Firstly, the ground states DFT calculations reveal that 2D MXenes, including $Cr_2VC_2F_2$, $Mo_2VC_2F_2$, $Mo_2VN_2F_2$, $Mo_3C_2F_2$ and $Mo_3N_2F_2$, have unusual ferrimagnetic order. Then, our real time time-dependent DFT (TDDFT) simulations on FiM MXenes have shown that laser pulses induce spin-selective charge transfer and further generate dramatic changes in the magnetic structure of these MXenes, including a magnetic order transition from FiM to transient FM at early time. The microscopic mechanism underpinning this ultrafast switching of magnetic order is governed by optically induced inter-site spin transfer (OISTR) effect. The results open new opportunities to manipulate the 2D magnetism by optical approaches

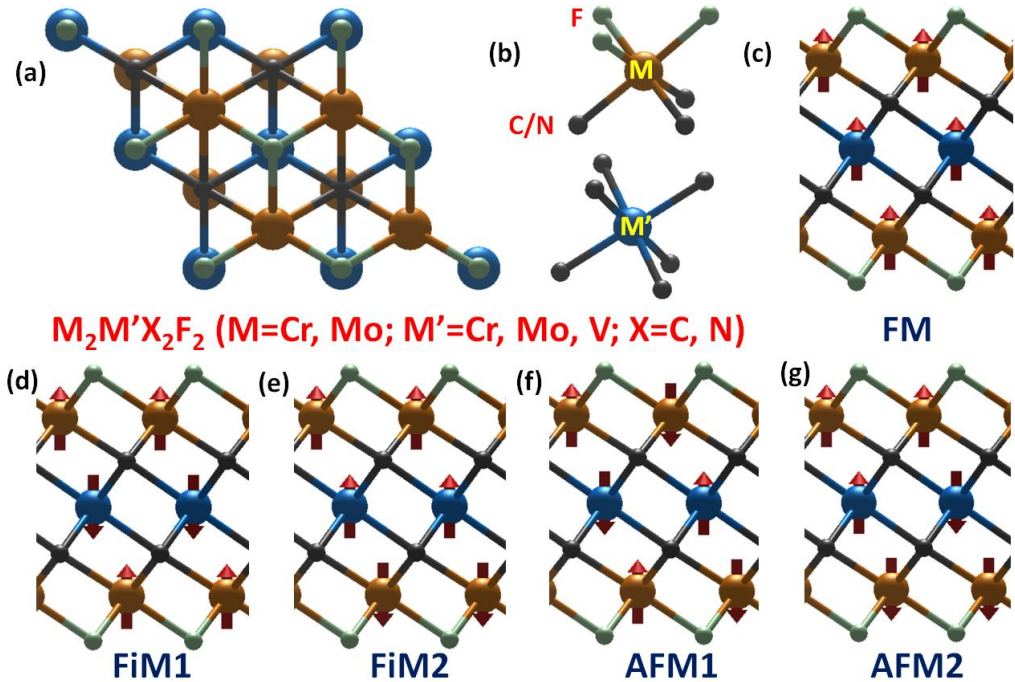

**Fig 1:** The top (a) $M_2M'X_2F_2$ (M=Cr, Mo; M'=Cr, Mo, V; X=C/N). (b) The ligand environment for M and M' is shown. Schematics of magnetic configurations for considered MXenes: (c) FM, (d) FiM1, (e) FiM2, (f) AFM1, (g) AFM2. The 2×1 supercells are employed for the total energy calculation.

## Results and discussions

The atomic structure MXenes-based 2D systems, including $Cr_2VC_2F_2$, $Mo_2VC_2F_2$, $Mo_2VN_2F_2$, $Mo_3C_2F_2$, and $Mo_3N_2F_2$ show in Figure 1. The M and M' localized in the distinctive octahedral environment, where M is bonded to six C/N atoms and M' atom, but M' is bonded to three C/N and three F atoms, respectively. The distinctive chemical environment renders the different local magnetic moment for M and M' atoms. Geometry optimization carried out at PBE+U level gives lattice constant and geometry parameters as summarized in Table S1. For $Mo_3N_2F_2$ MXene, the calculated constant lattice is in good agreement with previous calculations.[36] The calculated local magnetic moments for M and M' atoms are shown in Table 1. The M and M' atoms of MXenes predominantly contribute to the total magnetic moments while the neighboring X and F atoms have only a small contribution. To study the magnetic ground state structures of the $M_2M'X_2F_2$ MXenes, the collinear ferromagnetic (FM) and antiferromagnetic (AFM) and ferromagnetic (FiM) states were considered, respectively. To elucidate the spin configurations of these phases, we schematically plotted the arrangement of the local magnetic moments of M and M' atoms with colored arrows as shown in Figure 1c-g, which is considering a $2 \times 1$ supercell of MXenes. The lowest energy has been found for the FiM1 state and this has been set as reference energy. For instance, for $Cr_2VC_2F_2$ MXenes, the $E_{FM}$, $E_{FiM2}$, $E_{AFM1}$, and $E_{AFM2}$ states have energies 0.81, 0.25, 0.83, and 0.35 eV higher, respectively, than the FiM1 state. These results clearly demonstrate that the FiM1 state configuration for all considered MXenes is the magnetic ground state with high magnetic stability, similar to previously reported results for $Mo_3N_2F_2$ MXenes.[36] The magnetic anisotropy energy (MAE) of materials, which determines the orientation of the magnetization at low-temperature, is an important parameter for their applications in high-density storage or quantum spin processing. The non-collinear magnetic calculations were performed for magnetization along X[100], Y[010], and Z[001] directions for MXenes, which is summarized in Table S1. The results indicate that the easy axis of $Cr_2VC_2F_2$, $Mo_3C_2F_2$, and $Mo_3N_2F_2$ MXenes is along the [001] orientation, while easy axis of $Mo_2VN_2F_2$ and $Mo_2VC_2F_2$ is along the [100] and [010] orientation. The band structures with the orbital projection for MXenes in FiM state are reported in the supporting materials (see Figure S1). They all show the metallic feature. Interestingly, both $Mo_3N_2F_2$ and $Mo_2VN_2F_2$ are predicted to be half-metallic, which is a great potential for spintronic applications.

**Table1**: Calculated structural and magnetic characteristics of five different MXenes. [a]

| Structure | $E_{FM}$ | $E_{FiM1}$ | $E_{FiM2}$ | $E_{AFM1}$ | $E_{AFM1}$ | $\mu_M$ | $\mu_{M'}$ |
|---|---|---|---|---|---|---|---|
| $Cr_2VC_2F_2$ | 0.81 | 0 | 0.25 | 0.83 | 0.35 | 2.39 | -0.43 |
| $Mo_2VC_2F_2$ | 0.54 | 0 | 0.27 | 0.58 | 0.16 | 1.03 | -0.96 |
| $Mo_2VN_2F_2$ | 2.33 | 0 | 2.46 | 1.52 | 1.52 | 1.91 | -1.90 |
| $Mo_3C_2F_2$ | 0.96 | 0 | 1.07 | 0.89 | 1.18 | 1.39 | -0.25 |
| $Mo_3N_2F_2$ | 1.30 | 0 | 0.75 | 0.90 | 1.09 | 1.21 | -0.73 |

[a] Relative energies for FM ($E_{FM}$) , FiM1 ($E_{FiM1}$), FiM2 ($E_{FiM2}$), AFM1 ($E_{AFM1}$)  and AFM2 ($E_{AFM2}$) are shown in eV. Local magnetic moments of M ($\mu_M$) and M' ($\mu_{M'}$) atoms for $M_2M'X_2F_2$ MXenes are also given in $\mu_B$.

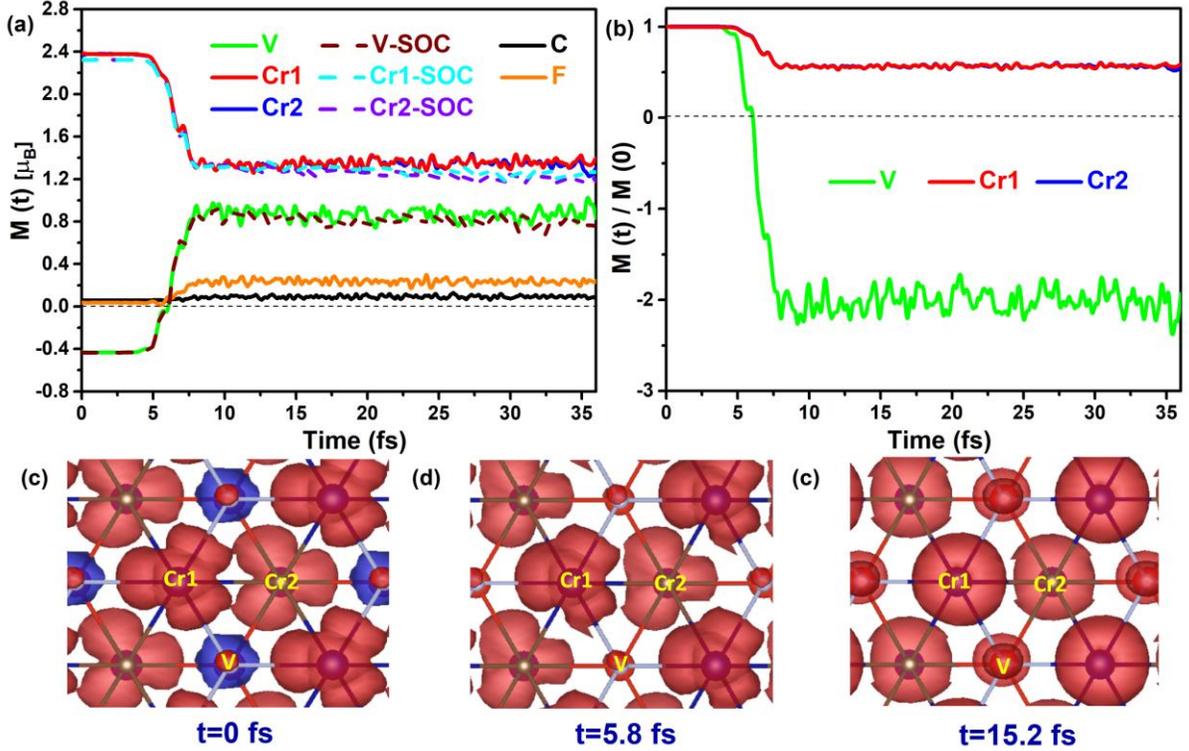

**Figure 2:** Ultrafast laser-induced ultrafast change of magnetic order. (a) Time evolution of local magnetic moment with (full lines) and without (dashed lines) spin-orbital coupling for Cr1, Cr2, V, C and F atoms, which at t = 0 fs (i.e., in the ground state) is ferrimegnets.  At ~6 fs, the V atom demagnetizes and then remagnetizes but with the reversed spins direction. The magnetic order thus switches from FiM to transient FM order. (b) The relative magnetization

dynamics for Cr1, Cr2 and V atoms. The snapshots of magnetization density at (c) t = 0 fs, (d) t=5.8 fs, and (e) t=15.2 fs. The iso-surface was set to be 0.012 e/$\text{Å}^3$

Next, we will explore the real-time evolution of the element resolved magnetism of the FiM MXenes monolayer induced by the electron dynamics under the influence of an external light with a selected frequency. We performed real-time TDDFT calculations, in which the linearly polarized (in-plane polarization) laser pulse with full width at half-maximum (FWHM) of 3.63 fs, photon energy of 1.63 eV and fluence of 34.2 mJ/$\text{cm}^2$ (the vector potential $\mathbf{A}$(t) of the laser pulse are shown in Figure S2(c)). We first examined the real-time spin dynamics of local magnetic moment of Cr and V atoms for $Cr_2VC_2F_2$ MXenes are shown in Figure 2a under the influence of a laser pulse. We can see that both the local moment Cr and V atoms change dramatically, while the global moment showed no change. The top and bottom Cr atoms (Cr1 and Cr2) demagnetize strongly from ~2.4 $\mu_B$ of initial magnetic moment to final ~1.4 $\mu_B$. Most interestingly, at ~6 fs (slightly before peak intensity of the laser pulse) the V atom demagnetizes and then remagnetizes (from ~ -0.4 $\mu_B$ to ~ 0.9 $\mu_B$) but with the reversed spin direction, thus leading to the FiM-FM magnetic order transition. After 10 fs, the $Cr_2VC_2F_2$ shows transient FM order for a long time (at least up to 100 fs, at which point our simulation ended). Under the influence of an external laser field, the relative local moments, i.e., the quantity (M(t)/M(0)) changes dramatically; the moment of Cr atoms show about 50% decrease, while the ultrafast loss of moment for V atoms and then increase (~200%) but with the reversed spin direction as shown in Figure 2b. We also simulate the real-time spin dynamics of $Cr_2VC_2F_2$ MXenes with spin-orbital coupling (SOC) as shown in Figure 2a. Because of the total spin moment will no longer a good quantum number, the spin-flip of photo-excited processes in $Cr_2VC_2F_2$ MXenes will be allowed. It is clear that that the effect of SOC will be not notable at least below 36 fs. The dynamics of the magnetization density as a function of time are shown in Figure 1d. The switching of the spin direction of V at t = 15.2 fs can also be seen in this plot.

The magnetic states transition from FiM to FM in $Cr_2VC_2F_2$ can be understood by optically induced inter-site spin transfer (OISTR) mechanism, which is found previously to dominate the early time spin dynamics of magnetic metal multilayers and the Heusler compounds.[25,26] The ground-state density of states (DOS) for Cr and V atoms shown in Figure 3a. Because the Cr and V are ferrimagnetic coupling, the majority spin is oppositely oriented in each atoms: spin-up is the majority in the Cr1 and Cr2 atoms, and spin-down is the majority in V atom. From DOS, the occupied Cr atoms are dominated by majority states, whereas there is considerable empty minority orbital of V around [1, 3] eV energy window. Such ground states electronic structure will enable efficient photo-driven spin selective charge transfer from the majority Cr to minority V atoms according to the OISTR mechanism as shown in Figure 3b, resulting in the loss of magnetic moment for Cr atoms and the gain of magnetic moment for V atoms, respectively. The OISTR mechanism in magnetic order transition of $Cr_2VC_2F_2$ MXenes is in agreement with the previously reported magnetic metal multilayers and Heusler alloys.[25,26] To obtain further analysis for the changing magnetic moment for $Cr_2VC_2F_2$ MXenes, we show time-dependent occupation changes ($\Delta n(t)$) of the V and Cr atoms, which is defined as the difference of the time-resolved occupation function $n(t)$ at time $t$ concerning the unexcited $Cr_2VC_2F_2$ MXenes, i.e., $\Delta n(t) = n(t) - n(0)$. As shown in Figure 3, the most notable transient change below the Fermi level occurs in the minority channel of Cr and minority channels of V atoms. One can see that photo-excitation results in a significant loss in the number of majority spin carriers as a function of time, which corresponds to the demagnetization for Cr atoms. These loss minority spin carriers are transferred from the Cr to the V sub-lattice, thus, leading to the enhanced magnetic moment for V atom, and further induced the direction reverse. It is also evidenced an obvious gain in the number of minority spin carriers of V (See Figure 3d). For other possible excitation, the $\Delta n(t)$ for the Cr minority and V majority spin channel is relatively small as shown in Figure S3, which does not significantly affect the total spin-transfer process between Cr and V atoms. The time-dependent change in the spin-down and up electrons for V and Cr are presented in Figure S4 relative to the ground-state for $Cr_2VC_2F_2$ MXenes. It is clear from Figure S4 that majority electrons of Cr are transferred to the minority of V atoms, which is consistent with the OISTR mechanism.

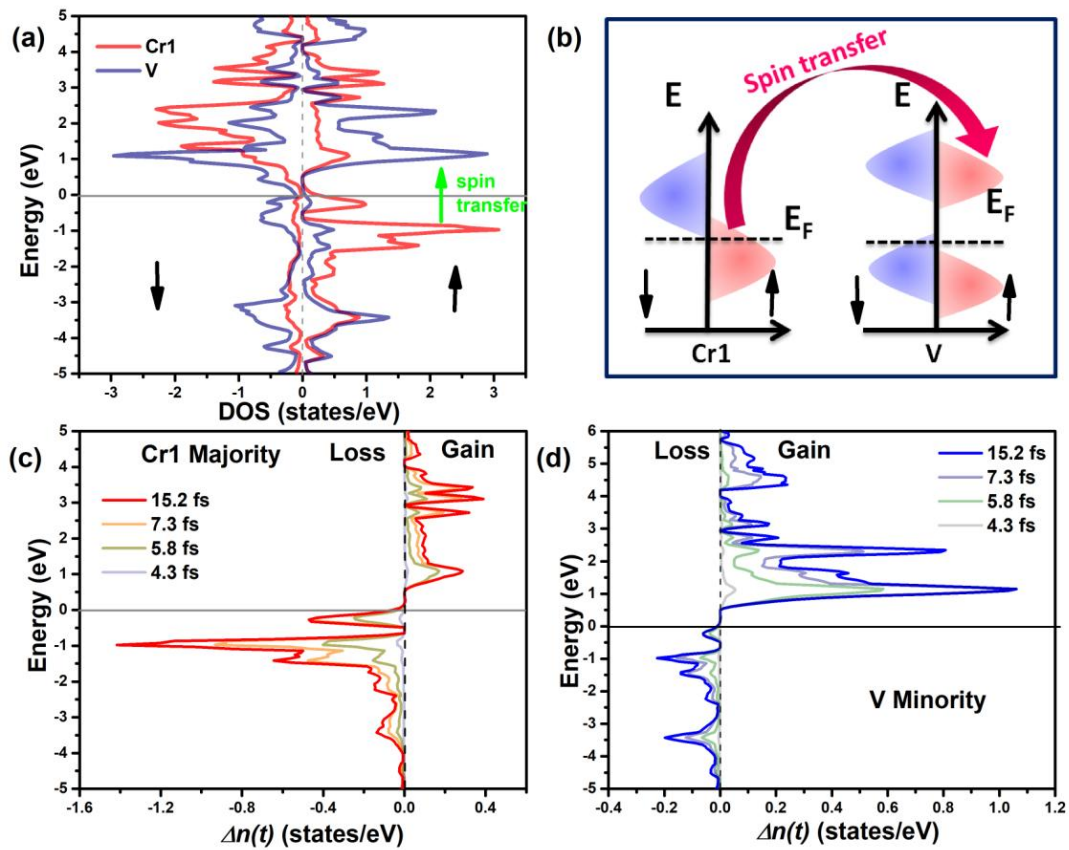

**Figure 3:** (a) Projected density of states (DOS) calculation for Cr1 and V in $Cr_2VC_2F_2$ MXenes. And, the favorable spin transfer from Cr majority to V minority is marked. (b) Schematic overview of the OISTR effect in $Cr_2VC_2F_2$. The optical excitation leads to an effective spin transfer from the occupied Cr majority into the V minority channel. TD-DFT calculations of the difference of the transient occupation compared with the unexcited case in the (c) Cr1 minority

channel and (d) V minority channel at selected time steps. For $\Delta$ n(t), a negative signal arises corresponding to a loss of minority electrons, while a simultaneous positive signal correlating to spin gain.

We now turn to explore the light response on magnetism for $Cr_2VC_2F_2$ MXenes with respect to the changing laser pulse parameter. Figure 4c has plotted the magnetization dynamics of Cr and V atoms in $Cr_2VC_2F_2$ MXenes under the influence of laser pulses of various frequencies (fixed FWHM) and FWHM (fixed frequency). Here, their time dependence of the external vector field is shown in Figure S2. It is clear that the demagnetization of Cr and V atoms is highly sensitive to the frequency of laser, indicating that the frequency of the pulse can also be used to tailor and optimize the process of photo-induced magnetic transition for FiM MXenes. For the higher energy in 1.09, 1.63, 2.18, and 2.72 eV, the local magnetic moment of Cr and V atoms are obviously demagnetized, whereas only relatively small loss of local magnetic moment is observed for energy in 0.54 eV. On the other hand, the demagnetization time will not strongly dependent on the FWHM. We can see V and Cr atoms are shown dramatic demagnetization for various FWHM values. However, the demagnetization process for various FWHM takes place after a time lag between the laser pulse and the maximum of the laser pulse (See Figure S2). Such a time lag effect between the laser pulse and the demagnetization in 2D MXene system also is similar to the Fe, Co, and Ni metal.[37]

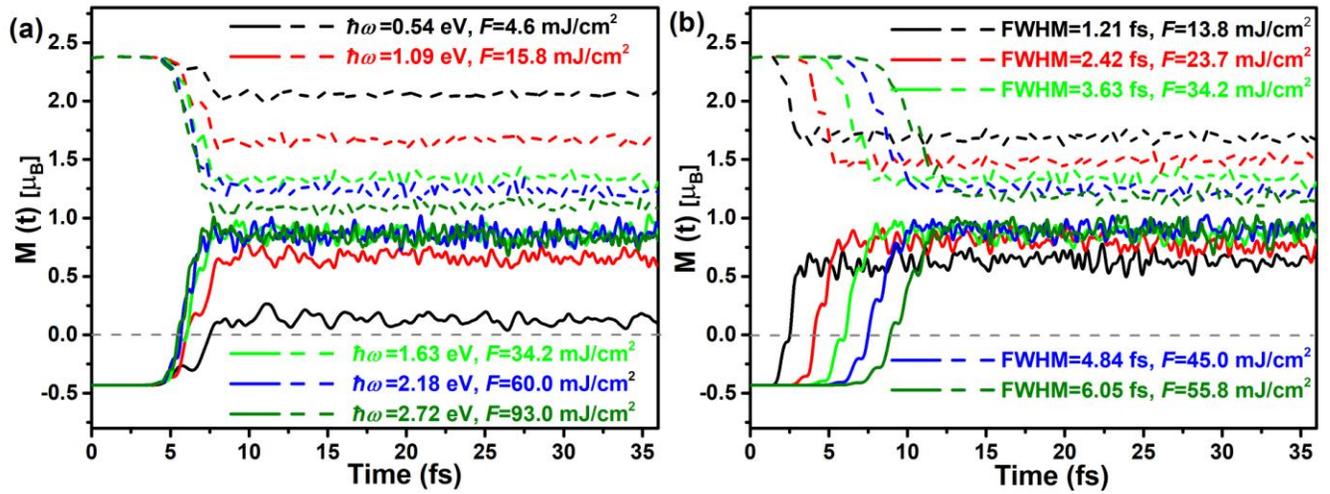

**Figure 4:** The spin dynamics of local magnetic moment of Cr1 and V atoms in $Cr_2VC_2F_2$ MXenes under the influence of seven different laser pulses given by (a) frequency/fluence, and (b) FWHM/fluence.

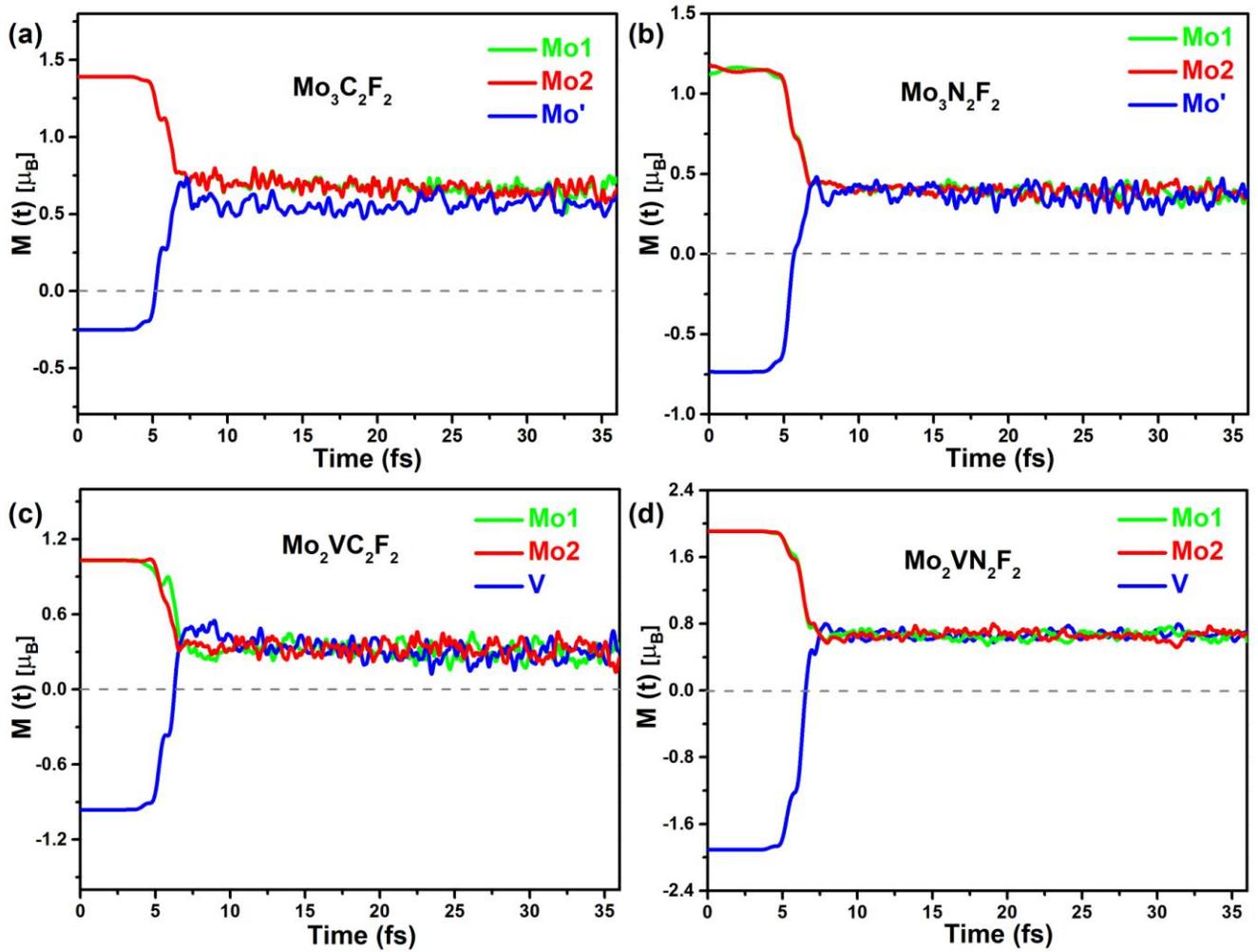

**Figure 5:** Time evolution of local magnetic moment of M and M' atoms for (a) $Mo_3C_2F_2$, (b) $Mo_3N_2F_2$, (c) $Mo_2VC_2F_2$ and (d) $Mo_2VN_2F_2$, respectively.

To probe the validity of optically induced magnetic order transition in other 2D MXenes materials, we have performed real-time TDDFT calculations of the laser induced change in magnetic structure for an extended set of 2D systems: $Mo_3C_2F_2$, $Mo_3N_2F_2$, $Mo_2VC_2F_2$, and $Mo_2VN_2F_2$, which have a FiM ground state. For purpose of companion, the same linearly polarized (in-plane polarization) laser pulse with FWHM of 3.63 fs, photon energy of 1.63 eV, and fluence of 34.2 mJ/cm$^2$ are employed as driven laser field. Under the influence of an external laser field, the local moment changes for these four materials dramatically change; the middle layered metal (M') demagnetize and then remagnetize with reverse spin direction, while the upper and lower metal (M) shown relatively small demagnetization as shown in Figure 5 and Figure S5. The results demonstrate that OISTR also dominates early-time spin dynamics in these FiM MXenes, which is the same mechanism as found in the case of $Cr_2VC_2F_2$ MXenes. It is also important to mention that linearly polarized light in the x-direction for present work was used. We find that changing polarization direction (e.g., y-direction) of this linearly polarized light does not affect the process of photo-induced magnetic order transition of MXenes. Note that our simulations just performed at early spin dynamics (up to 100 fs) following the laser pulse. This time-scale is completely dominated by direct optical manipulation without the rotation of atomic magnetic moments.[29] After this time, the phonon-mediated spin-flip processes will efficiently affect the magnetization of materials, and which can lead to further a reduction in magnetization; however, such processes are not included in our TDDFT simulations.

To explore how the optical manipulation of magnetism in 2D FM and AFM materials, we take $Mn_2CF_2$ and $Cr_2CF_2$ as an example to simulate their spin dynamics by TDDFT with the same laser parameter as $Cr_2VC_2F_2$ as shown in Figure S6. Both my calculation and previous reports indicate that the $Mn_2CF_2$ and $Cr_2CF_2$ have FM and symmetrical AFM order. From Figure S6, the $Mn_2CF_2$ show almost identical demagnetization of Mn atoms, result in no change in magnetic order. The AFM $Cr_2CF_2$ will symmetrical demagnetization because of spin transfer between majority and minority of Cr atoms will be identical in both spin channels. Our simulation in 2D FM and AFM materials is in agreement with the previously proposed demagnetization mechanism in bulk materials (e.g. NiO).[20] Thus, developing suitable 2D ferromagnetic systems will be crucial to explore ultrafast optical switch of magnetic order in 2D limit, which will trigger the further materials search and design for photo-induced magnetic states transition in 2D magnets as well as their van der Waals heterostructures. Experimentally, our work can be verified by ultrafast extreme ultraviolet (EUV) high harmonic pulses and time-resolved magnetic circular dichroism (MCD), which have been widely applied to explore the element-specific spin dynamics in multi-component magnetic systems.[19,29]

In conclusion, we performed DFT and real-time TDDFT calculations for ground-states properties and optical manipulation of magnetism in MXenes-based 2D system. Our DFT calculations show that $M_2M'C_2F_2$ MXenes, including $Cr_2VC_2F_2$, $Mo_2VC_2F_2$, $Mo_2VN_2F_2$, $Mo_3C_2F_2$, and $Mo_3N_2F_2$, have unusual ferrimagnetic order. Using TDDFT, we predicted that laser pulses can directly induce ultrafast spin-selective charge transfer between two magnetic sublattices (M and M') on a few femtoseconds, and further generate dramatic changes in the magnetic structure of these MXenes, including a magnetic order transition from ferrimagnetic (FiM) to transient ferromagnetic (FM). The spin transfer of magnetic sublattice in MXenes can further be tuned by laser parameter, which is achievable with current experimental techniques. The microscopic mechanisms for ultrafast switching of magnetic order are discussed based on optically induced inter-site spin transfer (OISTR) effect, which theoretically enables the ultrafast direct optical manipulation of the magnetic state in 2D magnets. Our results open new opportunities to manipulate the spin in 2D magnets as well as the potential applications in spintronics.

## Methods and computational details

The structure optimizations, magnetic ground states, and band structures for MXenes were performed using the Vienna *ab initio* simulation package (VASP)[38,39] within the generalized gradient approximation, using the Perdew-Burke–Ernzerhof (PBE) exchange-correlation functional.[40] Interactions between electrons and nuclei were described by the projector-augmented wave (PAW) method. The criteria of energy and atom force convergence were set to $10^{-6}$ eV and 0.001 eV/Å, respectively. A plane-wave kinetic energy cutoff of 500 eV was employed. The vacuum space of 15 Å along the MXenes normal was adopted for calculations on monolayers. The Brillouin zone (BZ) was sampled using $15 \times 15 \times 1$ Gamma-centered Monkhorst-Pack grids for the calculations of relaxation and electronic structures. To account for the energy of localized $3d$ orbitals of TM atoms properly, the Hubbard "U" correction is employed within the rotationally invariant DFT + U approach.[41] A correction of U = 3 eV for V, Cr, and Mo is employed based on the relevant previous reports.[11,12,14]

To identify the spin dynamics in these MXenes materials under the influence of ultrafast laser pulses, we have performed time-dependent density functional theory (TDDFT) calculations. The time evolving state functions ($\psi$) are calculated by solving the time dependent Kohn−Sham (KS) equation as follows:

$$i \frac{\partial \psi_j(\mathbf{r},t)}{\partial t} = \left[ \frac{1}{2} \left( -i\nabla + \frac{1}{c}\mathbf{A_{ext}}(t) \right)^2 + v_s(\mathbf{r},t) + \frac{1}{2c}\sigma \cdot \mathbf{B_s}(\mathbf{r},t) + \frac{1}{4c^2}\sigma \cdot (\nabla v_s(\mathbf{r},t) \times -i\nabla) \right] \psi_j(\mathbf{r},t) \quad (1)$$

where $\mathbf{A_{ext}}(t)$ and σ represent a vector potential and Pauli matrices. The KS effective potential $v_s(\mathbf{r},t) = v_{ext}(\mathbf{r},t) + v_H(\mathbf{r},t) + v_{xc}(\mathbf{r},t)$ can be decomposed into the external potential $v_{ext}$, the classical Hartree potential $v_H$, and the exchange-correlation (XC) potential $v_{xc}$, respectively. The KS magnetic field can be written as $\mathbf{B_s(r}, t) = \mathbf{B_{ext}(r}, t) + \mathbf{B_{xc}(r}, t)$, where $\mathbf{B_{ext}}$ and $\mathbf{B_{xc}}$ represent the magnetic field of the applied laser pulse plus possibly an additional magnetic field and XC magnetic field, respectively. The last term in Eq. (1) stand for the SOC. We only time propagate the electronic system while keeping the nuclei fixed.

With the $\psi$ s in hand a time-resolved DOS, shown in Figure 2, can be calculated using the following expression:

$$n_\sigma(E, t) = \sum_i \int_{BZ} d^3k \, \delta\big(E - \varepsilon_{i\vec{k}\sigma}\big) g_{i\vec{k}\sigma}(t) \qquad (2)$$

with

$$g_{i\vec{k}\sigma}(t) = \sum_i n_{i\vec{k}\sigma} \left| \int d^3r \, \psi_{i\vec{k}\sigma}(\vec{r}, t) \psi^*_{i\vec{k}\sigma}(\vec{r}, 0) \right|^2 \qquad (3)$$

here, spin-resolved and time-dependent $g_{i\vec{k}\sigma}(t)$ is calculated from the projection of the time-propagated orbitals $\psi_{i\vec{k}\sigma}(\vec{r}, t)$ onto the ground-state Kohn-Sham orbitals at $t = 0$. The dynamical evolution of the electronic structure for MXenes after photo-excitations also can be analyzed by the time-dependent changes of $n_\sigma(E, t)$.

We employed a fully non-collinear version of TDDFT by full-potential augmented plane-wave ELK code.[42] A regular mesh in k-space of $7 \times 7 \times 1$ is used, and a time step of $\Delta t = 0.1$ a.u. is employed for the real time TDDFT simulation. A smearing width of 0.027 eV is used. Laser pulses used in the present work are linearly polarized (in-plane polarization) with selected frequency. All calculations were performed using adiabatic local spin density approximation (ALSDA) with Hubbard U (ALSDA+U), with U=3 eV for Cr, V, and Mo atoms.

# Supporting Information "Optically driven ultrafast magnetic order transitions in two-dimensional ferrimagnets"

Junjie He,[1,*] Thomas Frauenheim[1,2,*]

[1]*Bremen Center for Computational Materials Science, University of Bremen, Am Fallturm 1, 2835, Bremen, Germany.*

[2]*Beijing Computational Science Research Center (CSRC), Beijing 100193 and Shenzhen Computational Science and Applied Research (CSAR) Institute, Shenzhen 518110, China.*

E-mail: junjie.he.phy@gmail.com; thomas.frauenheim@bccms.uni-bremen.de

Table S1: Calculated constant lattice and magnetic anisotropy of five different MXenes. The L and h is the lattice constant and height of MXenes in Å. The magnetic anisotropy between Y and Z ($E_{YZ}$) and X and Z ($E_{XZ}$) directions are in meV. The ES stand for the electronic structure.

| Structure | $L$ | $h$ | $E_{XZ}$ | $E_{YZ}$ | easy axis | ES |
|-----------|-----|-----|----------|----------|-----------|-----|
| $Cr_2VC_2F_2$ | 3.04 | 7.04 | -1.04 | -2.40 | [001] | Metal |
| $Mo_2VC_2F_2$ | 3.26 | 7.07 | -5.70 | 2.38 | [010] | Metal |
| $Mo_2VN_2F_2$ | 3.24 | 7.27 | 2.58 | 2.90 | [100] | Half-metal |
| $Mo_3C_2F_2$ | 3.29 | 7.10 | -42.11 | -43.66 | [001] | Metal |
| $Mo_3N_2F_2$ | 3.35 | 6.87 | -0.21 | -0.26 | [001] | Half-metal |

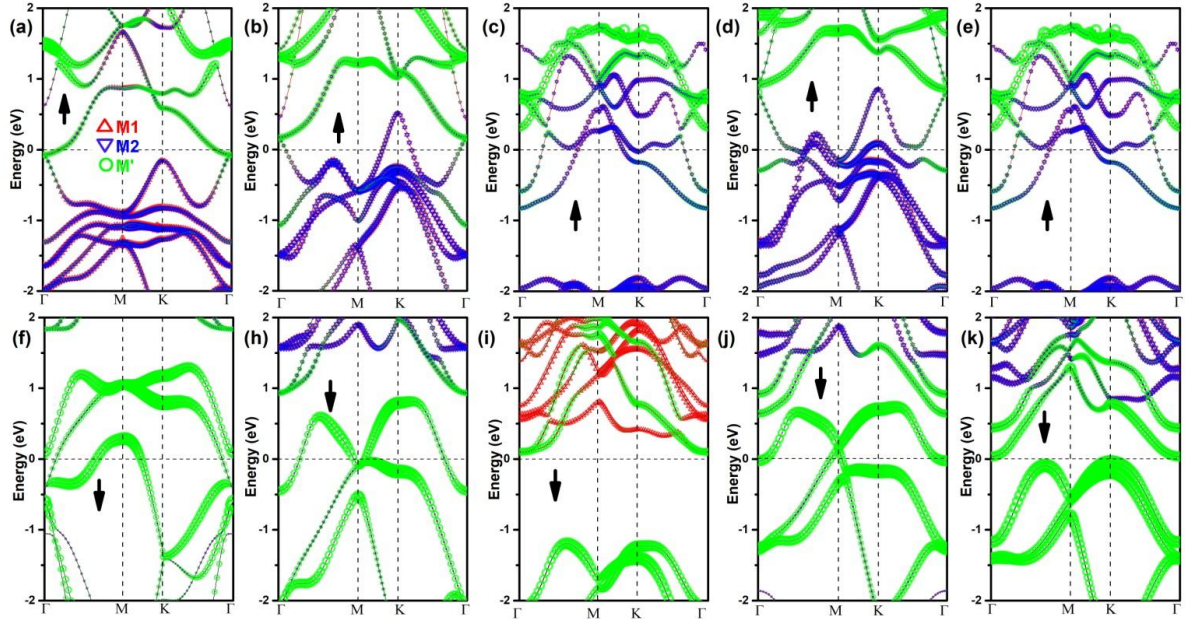

**Figure S1:** Band structure with orbital projection for spin up of (a) $Cr_2VC_2F_2$, (b) $Mo_3C_2F_2$, (c) $Mo_3N_2F_2$, (d) $Mo_2VC_2F_2$, (e) $Mo_2VN_2F_2$; Spin down of (f) $Cr_2VC_2F_2$, (g) $Mo_3C_2F_2$, (h) $Mo_3N_2F_2$, (i) $Mo_2VC_2F_2$, (g) $Mo_2VN_2F_2$.

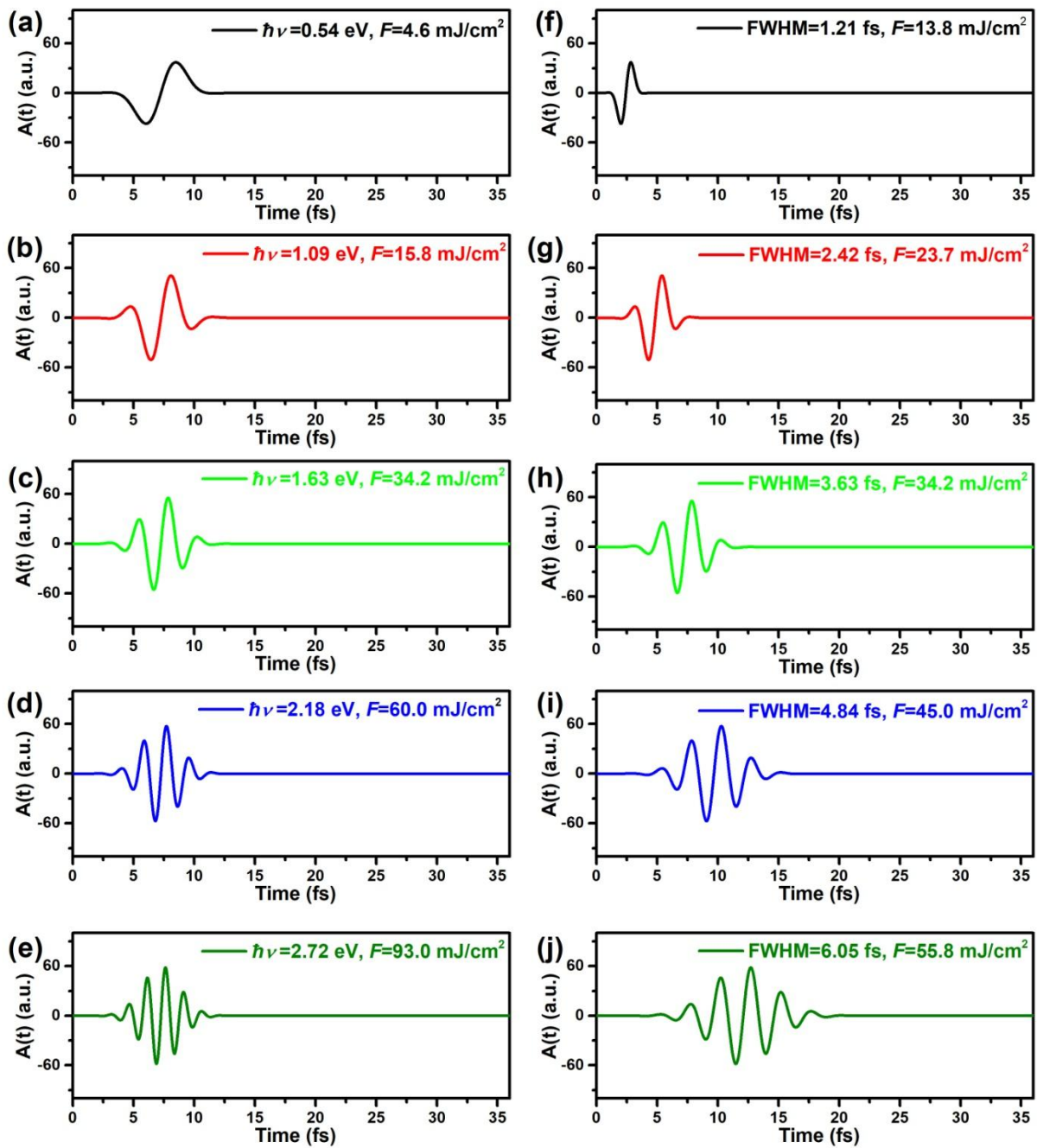

**Figure S2:** The applied pump laser pulse in $Cr_2VC_2F_2$. The energy and FWHM of laser was marked, respectively.

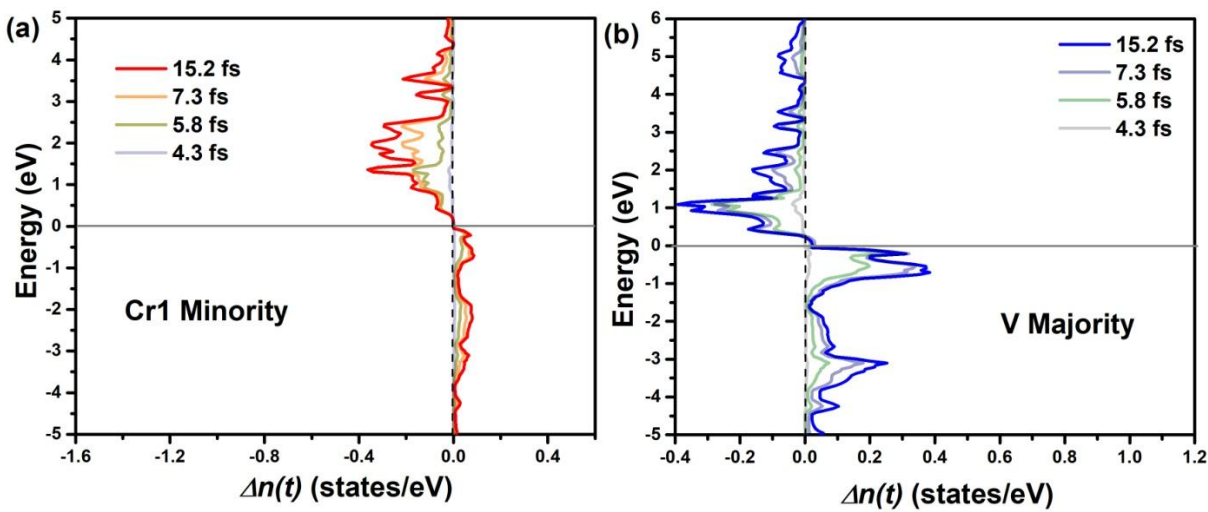

**Figure S3:** TDDFT calculations of the difference of the transient occupation compared with the unexcited case in the (c) Cr1 minority channel and (d) V minority channel at characteristic time steps. For Δn(t), a negative signal arises corresponding to a loss of minority electrons, while a simultaneous positive signal correlating to spin gain. The same x-axis scale in Δn(t) is selected with Figure 3cd

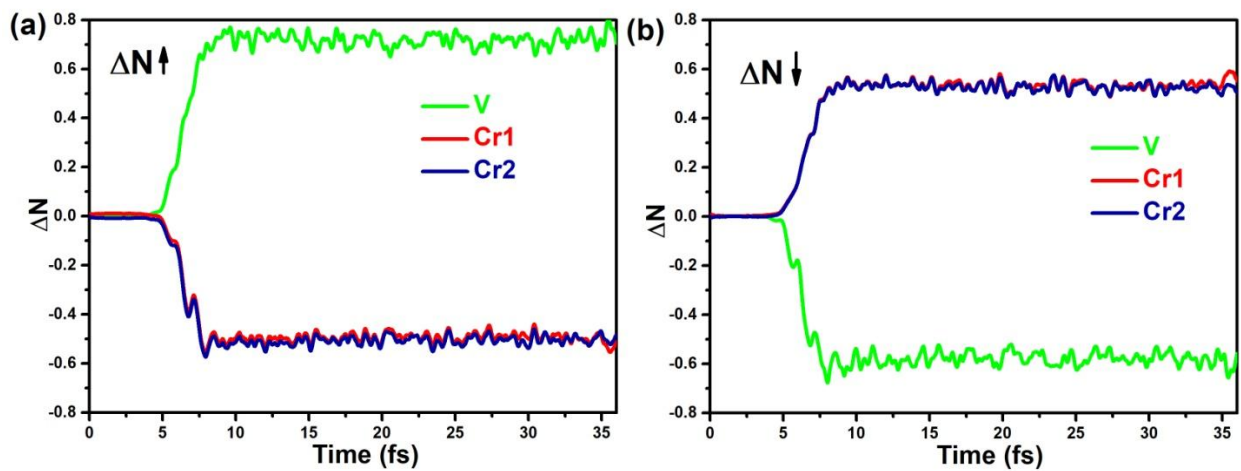

**Figure S4:** The time-dependent change in the (a) spin-up and (b) spin-down electrons on V, Cr1 and Cr2 atoms relative to the ground-state for $Cr_2VC_2F_2$.

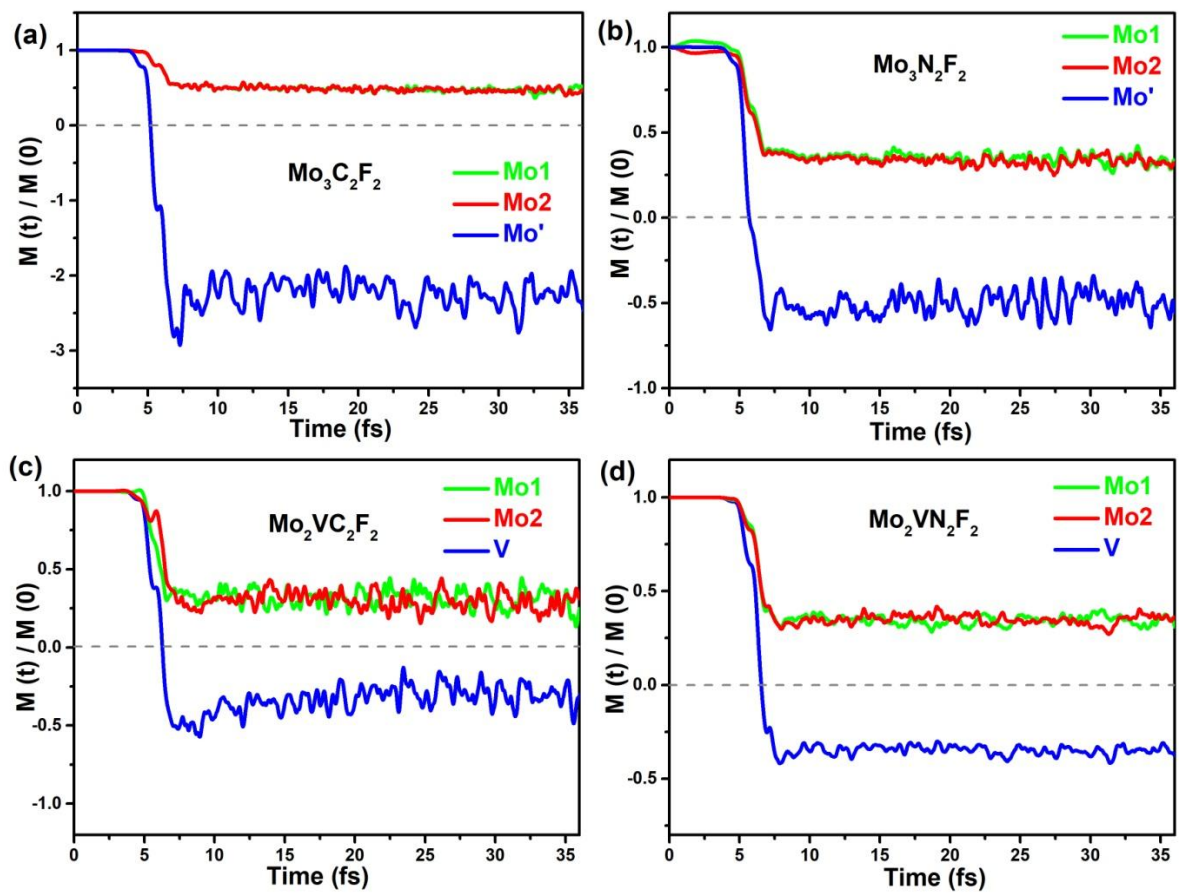

**Figure S5:** The relative local moment (i.e., the quantity ($M(t)/M(0)$) for (a) $Mo_3C_2F_2$, (b) $Mo_3N_2F_2$, (c) $Mo_2VC_2F_2$ and (d) $Mo_2VN_2F_2$.

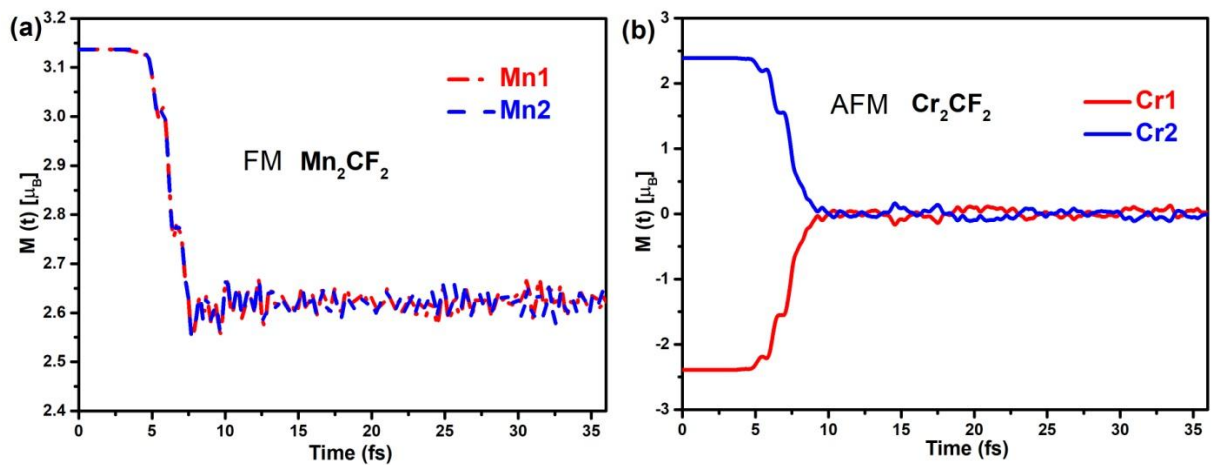

**Figure S6:** Photo-induced spin dynamics for (a) ferromagnetic MXenes $Mn_2CF_2$ and (b) antiferromagnetic MXenes $Cr_2CF_2$.